# The nonlinear physical aging-susceptibility of a glass forming melt

Kevin Moch,[1] Roland Böhmer,[1] and Catalin Gainaru[2]

[1] *Fakultät Physik, Technische Universität Dortmund, 44221 Dortmund, Germany*
[2] *Chemical Sciences Division, Oak Ridge National Laboratory, Oak Ridge, Tennessee 37831, USA*

A high-resolution, temperature oscillation-based probe of physical aging in complex systems is introduced. The Fourier analysis of the measured responses allows one to extract a high-order aging-related complex susceptibility that is not accessible via traditional temperature-jump and -ramp procedures. To demonstrate the potential of this oscillatory approach, we analyze the periodic time evolution of glycerol's structural relaxation using shear rheology as a vehicle. Thereby, we access up to the sixth harmonics and detect aging fingerprints within a resolution range of three orders of magnitude for temperature amplitudes of up to 4 K. The high-order aging coefficients and susceptibilities obtained for glycerol are described well by the Tool-Narayanaswami-Moynihan formalism.

Subjected to external perturbations such as provided by large mechanical loadings or thermal stresses, many materials modify their properties in a way that widely differ from those observed under small excitations. Founded on linear-response theory which describes the latter regime, one of the challenges in this area is to probe and predict the material response under aperiodic, step-like, or oscillatory large-scale excitations of any amplitude. A field of basic and practical relevance in which this approach is witnessing tremendous activity and consequently undergoing rapid progress is that of rheology. Here, Fourier transform (FT)[1,2] and related techniques[3] allowed one to characterize the dynamics of glass forming materials such as polymers in great detail. By virtue, *e.g.*, of the coupling of various relaxational modes within these materials, apart from the linear ($1\omega$) also the third ($3\omega$) and higher harmonics ($n\omega$ with $n > 300$) could occasionally be generated.[1] Often, the periodically imposed stresses or strains generate nonthermal states that eventually lead to wear, fatigue, and even complete material failure, a fascinating topic of its own.[4]

Cubic and higher-order susceptibility responses provide insights which can hardly be accessed in the linear-response regime, for instance regarding the dynamically heterogeneous nature and the cooperativity length of the structural relaxation of supercooled liquids,[5] regarding the effective ionic jump lengths and the emergence of anomalous Wien effects in ionic conductors,[6] as well as regarding morphological and dynamical rheological fingerprints that control the functionality of non-associating and associating polymers.[7,8] One of the challenges encountered in rheological, but also in nonlinear dielectric experiments,[9] is that the linear-response component is typically overwhelmingly large. Hence, clever schemes for its suppression have been devised.[5] An approach circumventing this problem is to exploit so-called cross experiments where, like, *e.g.*, in rheodielectric spectroscopy[10] the excitation and detection channels are well separated.

The field of thermal responses is of similar fundamental and technological relevance. Here, modulated calorimetry that implements excitation and detection *both* on the thermal channel, is usually carried out in the quasi-linear[11] regime. We are aware of only a few modulated cross-experiments that employ dielectric detection of the time dependence structural relaxation (in the quasi-linear regime) or of its thermally induced response.[12,13] The corresponding changes, called physical aging, are otherwise traditionally probed using temperature up- or down-jumps.[14] With the asymmetry[15] of the jump response being questioned,[16] and avoiding cumbersome experimental implementations of a temperature "step", in the present work we devise a cross experiment featuring thermal modulation and, in our case, rheological detection that taps the full FT potential for investigating nonequilibrium phenomena.

Analogous to the nonlinearity-probing FT rheology,[1] we are introducing "FT physical aging" spectroscopy with the goal of increasing the resolution framework in which temperature-related nonequilibrium phenomena can be investigated. Although this approach can be used for any complex system, in the present work we exploit it to monitor the periodic structural recovery for glycerol, a paradigmatic glass former. For such materials, FT physical aging not only provides access to high-order material individualities (similar to FT rheology), but can be used to test and potentially to discriminate various descriptions of structural recovery.[12,17,18,19]

Glycerol (from Sigma-Aldrich, stated purity > 99%) was investigated as received in an MCR502 rheometer from Anton-Paar. While oscillating the temperature, the complex shear response was probed in a 4 mm geometry using small strain amplitudes (typ. 0.03%) at a frequency $\nu_s$. The latter was chosen to be much larger (1) than the inverse molecular relaxation time $1/(2\pi\tau)$ and (2) than the sub-mHz frequency $\nu_T$ of the externally imposed sinusoidal temperature modulation $T(t) = T_b + \Delta T \sin(2\pi \nu_T t)$ with amplitude $\Delta T$. Base temperatures $T_b$ above as well as be-low glycerol's calorimetric glass transition temperature $T_g$ = 189 K[20] were chosen. As demonstrated below, the thermally well insulated rheometer oven ensured a highly accurate oscillatory temperature input.

The shear response will be discussed in terms of the loss tangent $\tan\delta$. Here, $\delta$ refers to the phase lag between applied strain and detected stress. Since both parts of the complex shear modulus $G^* = G' + iG''$ are proportional to the instantaneous shear modulus, $G_\infty(T)$, the loss tangent, $\tan\delta = G''/G'$, is insensitive to temperature variations of this quantity. The inset of Fig. 1 illustrates the frequency dependent $\log(\tan\delta)$ that we isothermally recorded for glycerol above and slightly below $T_g$. In the $T_b$ range from 193 to 187 K the $\tan\delta$ spectra follow power laws. Hence, at a fixed frequency $\nu_s$ the (vertical) $\log(\tan\delta)$ variation directly gauges the (horizontal) change in the instantaneous structural relaxation time $\tau$. The proportionality between $\log(\tan\delta)$ and $\log\tau$ is given by the slope $s$ of the lines in the inset of Fig. 1. For this simple relationship to hold, the oscillation amplitude $\Delta T$ must not be too large. As the inset of Fig. 1 reveals, $s$ may change when employing $\Delta T$ larger than currently used.

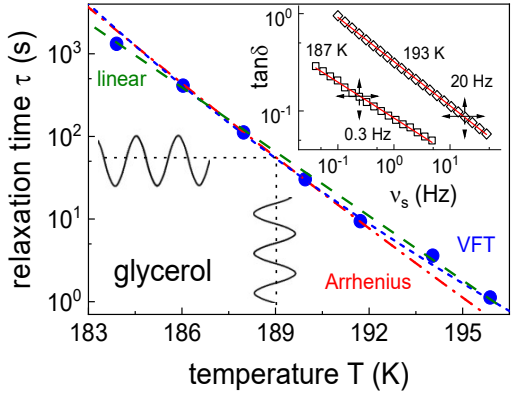

Fig. 1. In the covered temperature range glycerol's dielectric relaxation times[21], i.e., $\log\tau$ (blue dots), depend approximately (i) linearly on $T$ (green dashed line). With the goal to describe the same data points, the red dash-dotted line represents (ii) an Arrhenius law, $\log\tau \propto 1/T$, and the blue dotted line (iii) a Vogel-Fulcher-Tammann (VFT) law, $\log\tau \propto 1/(T - T_K)$ with a fit variable $T_K$.[21] The sinusoidal lines visualize how a temperature oscillation leads to a modulated $\log\tau$. The inset presents glycerol's frequency dependent shear loss tangent, $\log(\tan\delta)$. This quantity is probed in the stress-strain linear-response regime and shown here for two of the employed base temperatures. The red solid lines demonstrate the approximate linear relation between $\log(\tan\delta)$ and $\log\nu$. The vertical arrows indicate the frequencies used for the (shear-probed) temperature oscillations. The arrows schematically illustrate that a temperature-induced (horizontal) change in $\log\nu$ leads to a (vertical) $\tan\delta$ variation.

As the sine curves in Fig. 1 schematically illustrate, in the end it is important how the $T$ oscillation modulates $\log\tau$, irrespective of the particular quantity, in our case $\log(\tan\delta)$, or even of the experimental method, in our case shear rheology. Like in traditional step-like experiments,[22] the choice of the vehicle used to detect aging should not matter since it is not expected to alter the final outcome of the structural recovery.

Fig. 2(a) presents the rheological $\log(\tan\delta(t))$ induced by oscillating the temperature with frequency $\nu_T = 0.5$ mHz and amplitude $\Delta T = 3$ K about a base temperature $T_b = 193$ K. In this experiment the temperature does not cross glycerol's calorimetric $T_g$ and the appearance of $\log(\tan\delta(t))$ response is essentially symmetric with respect to the heating and cooling lobes. This is different for the tests carried out at 190 ± 3 K (for $\nu_T = 0.5$ mHz) and 187 ± 4 K (for $\nu_T = 0.33$ mHz), with the corresponding results shown in Fig. 2(b) and (c), respectively. The cycling about $T_b = 187$ K mostly explores the range below $T_g$, i.e., the glassy state where physical aging clearly affects the temperature modulated shear signal. Considering the asymmetry of the latter, not only odd, but also even[23] harmonics can be expected to occur in the Fourier analysis of the $\log(\tan\delta(t))$ response.

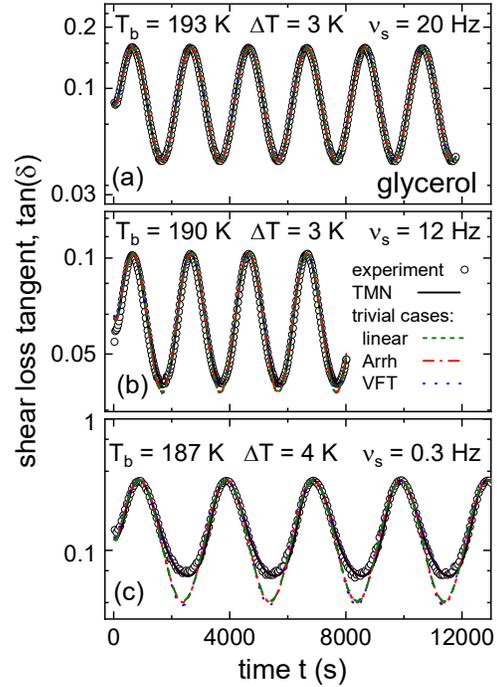

Fig. 2. Shear loss tangent (open symbols) probed for glycerol at the indicated frequencies $\nu_T$ using temperature oscillations with amplitudes $\Delta T$ about the base temperatures $T_b$. The black solid lines are calculations based on the Tool-Narayanaswami-Moynihan model, cf. Eqs. (2) and (3). The other lines reflect "trivial" equilibrium expectations defined by assuming that fictive and thermodynamic temperature agree. The corresponding lines refer to the linear, Arrhenius, and VFT dependences of $\log\tau(T)$, see Fig. 1.

The detection limit of the harmonics in the output channel, $\log(\tan\delta(t))$, is set by deviations from the (ideally) sinusoidal temperature input. Using the present setup, the green stars in Fig. 3 show that for their relative intensity we find $\theta_n/\theta_1 \leq 10^{-3}$. Here the $n^{\text{th}}$-order thermal intensity $\theta_n = |X_n^*|$ is defined via

$$X_n^* = X_n' + iX_n'' = \frac{2}{T_p}\int_0^{T_p} X(t)\,e^{2\pi i \nu_T n t}\,dt \qquad (1)$$

with $|X_n^*| = \sqrt{(X_n')^2 + (X_n'')^2}$ and $X(t) = T(t)$. The length of the detection period is designated $T_p$. As we will demonstrate below, this very low input distortion level enables the reliable detection of up to the sixth harmonic in $\log(\tan\delta(t))$. Based on the shear responses presented in Fig. 2, Fig. 3 summarizes the $n^{\text{th}}$-order Fourier intensities $I_n$, defined in analogy to $\theta_n$, but now with $X(t) = \log(\tan\delta(t))$. This choice for $X(t)$ reflects the fact that to an excellent approximation $\log(\tan\delta)$, and not for instance $\tan\delta$, is proportional to $\log\tau$.

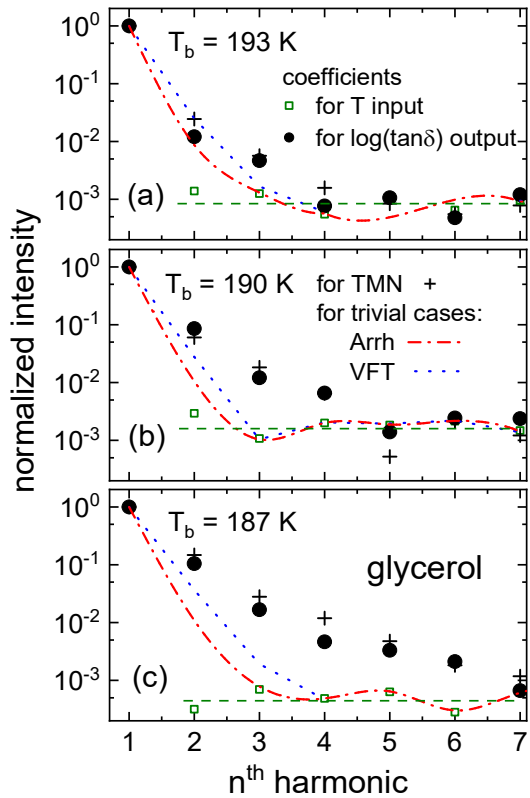

Fig. 3. Normalized (with respect to the intensity of the fundamental mode) $n^{\text{th}}$-order Fourier coefficients obtained for temperature oscillation signals generated at (a) 193 K with $\Delta T = 3$ K, (b) 190 K with $\Delta T = 3$ K, and (c) 187 K with $\Delta T = 4$ K: These are the intensity $I_n/I_1$ of the measured output, $\log(\tan\delta) \propto \log\tau$, (circles), of the TMN model (plusses), and of the "trivial" $\log(\tan\delta)$ calculations[27] (for simplicity represented as curved lines). Clearly, the higher-order output signals (circles) grow much above the "trivial" signals as the base temperature is lowered. The normalized intensity $\theta_n/\theta_1$ of the temperature input (open squares) sets the resolution limit for the experimentally determined output coefficients as also highlighted by the vertical lines.

At $T_b = 193$ K, that is when $T(t)$ never crosses $T_g$ for the chosen $\Delta T$, the appearance of the weak second and third harmonics in Fig. 3(a) reveals that $\log\tau$ obeys Arrhenius or VFT rather than linear temperature dependences [the latter implies similar normalized Fourier coefficients for the $\log\tau(t)$ output and the $T(t)$ input]. Considering that deviations from $\log\tau \propto T$ are fairly indiscernible from Fig. 1, the presence of "equilibrium" $2\omega$ and $3\omega$ components attests to the high sensitivity of the present experimental approach.

While for $T_b = 193$ K the harmonics with $n \geq 4$ are all buried in the noise, this changes for $T_b = 190$ K where $I_4$ is already nonzero, and even more so for $T_b = 187$ K, where the sixth harmonic is still significant.

In order to assess which contributions to $I_n$ are "trivial" in the sense alluded to above, it is best to rely on a well-tested concept for modeling (step- and ramp-induced) aging phenomena. One such approach is that named after Tool, Narayanaswami, and Moynihan (TNM). Within the premises of this formalism, the time evolution of $\tau$ is given by

$$\tau(t'') = A\exp\left[\frac{xh}{RT(t'')} + \frac{(1-x)h}{RT_f(t'')}\right], \qquad (2)$$

with $R$ the ideal gas constant, $A$ and $h$ the Arrhenius parameters describing the equilibrium $T$ evolution of $\tau$ in the temperature range of interest, $\tau = A\exp(h/RT)$. The sinusoidal input is represented by $T(t'')$, $x$ is the so-called nonlinearity parameter, and $T_f$ the fictive temperature which by definition is the "temperature at which the corresponding liquid structure and properties are frozen in upon cooling".[24] The time variation of $T_f$ can be deduced from

$$T_f(t) = T_0 + \int_{t_0}^{t}\left\{1 - \exp\left[-\left(\int_{t'}^{t}\frac{dt''}{\tau(t'')}\right)^\beta\right]\right\}q(t')dt', \qquad (3)$$

with $T_0 = T_b$ denoting the initial temperature at time $t \to t_0$, the exponent $\beta$ a stretch exponent, and $q(t') = dT(t')/dt'$ a rate. For the present analyses we use the effective equilibrium parameters $A = 1.7\times 10^{-54}$ s and $h = 200$ kJ/mol[25] and the reported TNM parameters $x = 0.29$ and $\beta = 0.51$.[26]

To assess the "trivial" $n^{\text{th}}$-order weights to the nonlinear contribution, Fig. 2 includes $\log(\tan\delta(t))$ estimates with $T_f$ in Eq. (3) set equal to the thermodynamic temperature $T$. This condition suppresses any dependence on the nonlinearity parameter $x$, as Eq. (2) reduces to an Arrhenius law that approximates the equilibrium temperature dependence of the relaxation time. Fig. 3 shows that for $T_b = 193$ K the resulting "trivial" intensities corresponding to the Arrhenius and VFT approximations match the experimental $I_n/I_1$ results well.[27] However, for $T_b = 190$ and 187 K the experimentally determined intensities are *significantly larger* than the calculated background nonlinearities.

Exploiting the proportionality between $\log\tau$ and $\log(\tan\delta)$, see the inset of Fig. 1, we performed TNM calculations also for the nontrivial case. This means that we drop the condition for $T_f$ and then solve Eq. (3) self-consistently for this quantity. Fig. 2 shows that the resulting predictions for $\log(\tan\delta(t))$ closely trace the experimental data, thereby providing additional confidence in the parameters used for the TNM analysis. The higher-order intensities displayed in Fig. 3(b) and (c) allow for a more sensitive check of the oscillatory aging approach: The calculated intensities furnish a more or less quantitative description of all of the measured intensities. Thus, our results demonstrate that the present method can be used to access feeble nonlinear fingerprints of relaxation processes, similar to what is achieved using FT rheology.[1,2]

Exploiting the amplitude and phase shift of the output, see Fig. 2, with respect with the oscillatory input, one can access the complex nonlinear aging-susceptibility $\chi^*_{Am}$ defined by the coefficients given in

$$\frac{\log[\tau/\tau(T_0)]}{\Delta T} = \sum_{m=1}^{\infty} [\chi'_{Am}\cos(2\pi m\nu_T t) + \chi''_{Am}\sin(2\pi m\nu_T t)]. \quad (4)$$

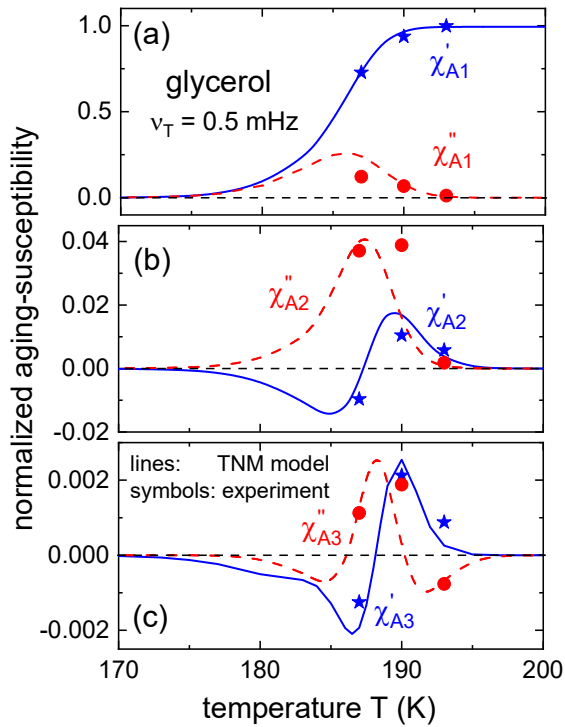

Fig. 4. The experimental real (stars) and imaginary (dots) parts of the currently introduced (a) first, (b) second, and (c) third-order aging-susceptibilities of glycerol are compared with the corresponding TNM predictions (represented as lines). The normalization refers to the high-temperature limit of $\chi'_{A1}$.

Unlike for most other susceptibilities which tacitly consider perturbations with respect to *zero* field, in Eq. (4) explicit reference to the (equilibrium relaxation time at the) initial temperature $T_0$ is necessary. Furthermore, it has to be realized that Eq. (4) represents an effective aging-susceptibility since, like the coefficients shown in Fig. 3(a), it features also "trivial" contributions and since its numerator does not refer to an extensive variable.

Fig. 4 reveals an acceptable agreement between real and imaginary parts of $\chi^*_{Am}$ accessed experimentally for $m\omega = 1\omega$, $2\omega$, and $3\omega$ and those generated using the TNM approach. The spectral features displayed in Fig. 4 resemble those displayed by, e.g., nonlinear dielectric susceptibility experiments.[28,29,30] This suggests that similar to dielectric and rheological first-order nonlinearities (i.e., $\chi_{13}$),[30,31,32,33] related quantities may be accessed in aging-susceptibility experiments performed at constant $\nu_T$ by varying $\Delta T$.

Systematically increasing the oscillation amplitudes $\Delta T$ from mK to several K will open the exciting possibility to experimentally relate structural fluctuations to structural recovery, which are fundamental aspects of the glass transition phenomenon. Furthermore, complementing the concept of fictive field[34] (instead of fictive temperature), the introduced susceptibility facilitates a generalized description of nonlinear dielectrically, rheologically, and thermally induced responses of glass forming materials.

To conclude, this letter adopts the Fourier transform approach used in frequency domain high-field dissipation studies, to access high-order nonlinearities associated with physical aging of glass forming materials. From an experimental standpoint, the employed temperature oscillation protocol precludes some pitfalls of traditional isothermal aging tests, such as finite rate and overshoots/undershoots of intended step-like temperature changes and the partial aging occurring during thermal equilibration. As demonstrated for glycerol by means of shear rheology, oscillatory aging experiments provide access to high-order coefficients of time-dependent relaxation times, unraveling up to its 6th harmonic. These high-resolution aging results could be described well using the TNM approach, and they can be further used to test or to develop alternative phenomenological or theoretical aging descriptions. Additionally, we introduced a complex nonlinear aging-susceptibility that captures the sample-specific dynamical changes induced by thermal perturbations. Thus, the present work supplies a powerful tool for the study of nonequilibrium phenomena in condensed matter and opens new venues for nonlinear probes in a multidimensional time-temperature space.

Financial support by the Deutsche Forschungsgemeinschaft Grant No. 461147152 is gratefully acknowledged.